\documentclass[aps,preprint,preprintnumbers,amsmath,amssymb,nofootinbib]{revtex4}
\usepackage{amssymb}
\usepackage{amsmath}
\usepackage{bm}
\usepackage{graphicx}
\usepackage{epsfig}
\begin{document}
\title{Quantum origin of pre-big-bang collapse from Induced Matter theory of gravity.}
\author{$^{1,2}$ Mauricio Bellini \footnote{ E-mail address:
mbellini@mdp.edu.ar, mbellini@conicet.gov.ar} } \vskip .2cm
\address{$^1$ Departamento de F\'isica, Facultad de Ciencias Exactas y
Naturales, Universidad Nacional de Mar del Plata, Funes 3350,
C.P. 7600, Mar del Plata, Argentina.\\  \\
$^2$ Instituto de Investigaciones F\'{\i}sicas de Mar del Plata (IFIMAR), \\
Consejo Nacional de Investigaciones Cient\'ificas y T\'ecnicas
(CONICET), Argentina.}

\begin{abstract}
We revisit a collapsing pre-big-bang model of the universe to
study with detail the non-perturbative quantum dynamics of the
dispersal scalar field whose dynamics becomes from the dynamical
foliation of test massless scalar field $\varphi$ on a 5D
Riemann-flat metric, such that the extra space-like coordinate is
noncompact. The important result here obtained is that the
evolution of the system, which is described thorough the equation
of state has the unique origin in the quantum contributions of the
effective 4D scalar field $\bar{\varphi}$.
\end{abstract}
\maketitle

\section{Introduction}

The five-dimensional model is the simplest extension of General
Relativity (GR), and is widely regarded as the low-energy limit of
models with higher dimensions (such as 10D supersymmetry and 11D
supergravity). Modern versions of 5D GR abandon the cylinder and
compactification conditions used in original Kaluza-Klein (KK)
theories, which caused problems with the cosmological constant and
the masses of particles, and consider a large extra dimension. In
particular, the Induced Matter Theory (IMT) is based on the
assumption that ordinary matter and physical fields that we can
observe in our 4D universe can be geometrically induced from a 5D
Ricci-flat metric with a space-like noncompact extra dimension on
which we define a physical vacuum\cite{IMT,imt1}. The
Campbell-Magaard
theorem\cite{campbell,c1,campbellb,campbellc,campbelld} serves as
a ladder to go between manifolds whose dimensionality differs by
one. This theorem, which is valid in any number of dimensions,
implies that every solution of the 4D Einstein equations with
arbitrary energy momentum tensor can be embedded, at least
locally, in a solution of the 5D Einstein field equations in
vacuum. Because of this, the stress-energy may be a 4D
manifestation of the embedding geometry. Physically, the
background metric there employed describes a 5D extension of an
usual de Sitter spacetime.  By making a static foliation on the
space-like extra coordinate, it is possible to obtain an effective
4D universe that suffered an exponential accelerated expansion
driven by a scalar (inflaton) field with an equation of state
close to a vacuum dominated one\cite{ua,ua1,LB,B}. The most
conservative assumption is that the energy density $\rho =P/\omega
$ is due to a cosmological parameter which is constant and the
equation of state is given by a constant $\omega =-1$, describing
a vacuum dominated universe with pressure $P$ and energy density
$\rho$. On the other hand, exists a kind of exotic fluids that may
be framed in theories with matter fields that violate the weak
energy condition\cite{u}, such that $\omega <-1$. These models
were called phantom cosmologies, and their study represents a
currently active area of research in theoretical
cosmology\cite{uu,uu1}.

On the other hand, the spherically symmetric collapse of a
massless scalar field has been of much interest towards
understanding the dynamical evolutions in general relativity. A
remarkable finding of some numerical investigations is the
demonstration of criticality in gravitational collapse.
Specifically, it was found that for a range of values of the
parameter characterizing the solution, black hole forms and there
was a critical value of the parameter beyond which the solutions
are such that the scalar field disperses without forming any black
hole. However, this result has been obtained mainly through
numerical studies and a proper theoretical understanding of this
phenomenon is still lacking (see e.g. \cite{3} and the references
therein). In order to study the dynamics of a massless scalar
field $\varphi$ on a 5D vacuum, we consider the canonical metric
\begin{equation}\label{m}
dS^2 = g_{\mu\nu} (y^{\sigma},\psi) d y^{\mu} dy^{\nu} - d\psi^2.
\end{equation}
Here the 5D coordinates are orthogonal: $y \equiv
\{y^a\}$\footnote{Greek letters run from $0$ to $3$, and latin
letters run from $0$ to $4$.}. The geodesic equations for a
relativistic observer are
\begin{equation}\label{geo}
\frac{dU^a}{dS} + \Gamma^a_{bc}\,\,U^b U^c =0,
\end{equation}
where $U^a = {dy^a\over dS}$ are the velocities and
$\Gamma^a_{bc}$ are the connections of (\ref{m}). Now we consider
a parametrization $\psi(x^{\alpha})$, where $x\equiv
\{x^{\alpha}\}$ are an orthogonal system of coordinates, such that
the effective line element (\ref{m}), now can be written as
\begin{equation}\label{met1}
dS^2 = h_{\alpha\beta}\,dx^{\alpha} dx^{\beta}.
\end{equation}
It is very important to notice that $S$ will be an invariant, so
that derivatives with respect to $S$ will be the same on 5D or 4D.
In other words, in this paper we shall consider spacetime lengths
that remain unaltered when we move on an effective 4D spacetime.

\subsection{Einstein equations for dynamical foliations from a 5D vacuum state}

Now we consider the Einstein equations on the 5D canonical metric
like (\ref{m})
\begin{equation}
G_{ab} = -8\pi G\, T_{ab},
\end{equation}
where the Einstein tensor is given by $G_{ab} = {\cal R}_{ab} -
\frac{1}{2} g_{ab} {\cal R}$ and ${\cal R}_{ab}$ is the Ricci
tensor, such that the scalar of curvature is ${\cal R}= g_{ab}
{\cal R}^{ab}$. Because we are considering a 5D Ricci-flat metric,
the Einstein tensor and the Ricci scalar will be null. Using the
transformations previously introduced, we obtain that
\begin{equation}\label{7}
\bar{G}_{\alpha\beta} = \bar{{\cal R}}_{\alpha\beta} - \frac{1}{2}
h_{\alpha\beta} \bar{{\cal R}} = -8\pi G \,\bar{T}_{\alpha\beta},
\end{equation}
where we have used respectively the transformations
\begin{eqnarray}
&& \bar{{\cal R}}_{\alpha\beta} =  e^a_{\,\,\alpha} e^b_{\,\,\beta} \,{\cal R}_{ab}, \\
&& \bar{{\cal R}} = h_{\alpha\beta} \bar{{\cal R}}^{\alpha\beta}, \\
&& \bar{T}_{\alpha\beta} = e^a_{\,\,\alpha} e^b_{\,\,\beta}
\,T_{ab},
\end{eqnarray}
for the effective 4D Ricci tensor, the scalar of curvature and the
energy momentum tensor.

\subsection{Energy-Momentum tensor}

We consider a quantum massless scalar field $\varphi(y^a)$ on the
metric (\ref{m}). In order to make a complete description for the
dynamics of the scalar field, we shall consider its energy
momentum tensor. In order to describe a true 5D physical vacuum we
shall consider that the field is massless and there is absence of
interaction on the 5D Ricci-flat manifold, so that
\begin{equation}
T^a_{\,\,b} = \Pi^a \Pi_b - g^a_{\,\,b} \,{\cal
L}\left[\varphi,\varphi_{,c}\right],
\end{equation}
where ${\cal L}\left[\varphi,\varphi_{,c}\right] = {1\over 2}
\varphi^a\varphi_a$ is the lagrangian density for a free and
massless scalar field on (\ref{m}) and the canonical momentum is
$\Pi^a = {\partial {\cal L}\over
\partial \varphi_{,a}}$. Notice that we are not considering interactions on the 5D vacuum,
because it is related to a physical vacuum in the sense that the
Einstein tensor is zero: $G^a_b =0$.

\subsection{Dynamics of the scalar field for a dynamical foliation}

We are interested to study how is the effective 4D dynamics
obtained from a dynamic foliation of a 5D Ricci-flat canonical
metric. We consider a classical massless scalar field
$\varphi(y^a)$ on the metric (\ref{m}). The effective 4D energy
momentum tensor will be
\begin{equation}
\bar{T}_{\alpha\beta} = \left.e^a_{\,\,\alpha} \, e^b_{\,\,\beta}
\,T_{ab}\right|_{\psi(x^{\alpha})}.
\end{equation}
In other words, using the fact that ${\cal L}$ is an invariant it
is easy to demonstrate that
\begin{equation}
\bar{T}^{\alpha}_{\,\,\beta} = \bar{\Pi}^{\alpha}\,
\bar{\Pi}_{\beta} - h^{\alpha}_{\,\,\beta} {\cal L},
\end{equation}
where ${\cal L}$ is an invariant of the theory: $ {\cal L} =
\frac{1}{2} \varphi^{,a} \varphi_{,a} = \frac{1}{2}
\left(e^a_{\alpha} \bar{\varphi}^{,\alpha}\right)
\left(\bar{e}^{\beta}_a \bar{\varphi}_{,\beta}\right)$. The
equation of motion for the scalar field $\bar\varphi$ becomes from
$\bar{\nabla}_{\alpha} \bar{T}^{\alpha}_{\beta} =0$\footnote{Here,
$\bar{\nabla}_{\alpha}$ denotes the covariant derivative on the
effective 4D hypersurface, with respect to the Christoffel
connections $\bar{\Gamma}^{\alpha}_{\beta\gamma}$.}, so that one
obtains
\begin{equation}\label{14}
h^{\mu\nu} \bar{\nabla}_{\nu} \bar{\varphi}_{,\mu}=0,
\end{equation}
that describes the dynamics of $\bar{\varphi}(x^{\alpha})$ on the
effective 4D hypersurface (\ref{met1}). Notice that in the
dynamics of $\bar\varphi$, which is described by eq. (\ref{14}),
it is absent any kind of interaction. This is because the
dynamical foliations as we are studied in this letter describe a
dispersal system\cite{Bh}.

In a previous letter\cite{an} we have studied the gravitational
collapse of the universe which is driven by a massless dispersal
scalar field. The system was studied from a 5D Riemann-flat
canonical metric, on which we make a dynamical foliation on the
extra space-like dimension. The asymptotic universe there
obtained, which is absent of singularities, results to be finite
in size and energy density, which tends to zero for asymptotic
large times, so that the asymptotic equation of state becomes
$\left.\omega\right|_{t\rightarrow \infty} \rightarrow -\infty$.
This is because the pressure is negative (opposes the collapse)
along all the contraction and its asymptotic value tends to zero,
but more slowly than does the energy density. In this letter we
shall revisit a collapsing system, but from a different 5D metric,
with the aim to study with detail the non-perturbative quantum
dynamics of the dispersal scalar field.

\section{An example: pre Big Bang collapsing universe}

We consider the 5D canonical extended de Sitter Riemann-flat
metric\cite{PdL}
\begin{equation}\label{m1}
dS^2 = \left(\frac{\psi}{\psi_0}\right)^2 \left[ dt^2 -
e^{-2\psi^{-1}_0 t } dr^2 \right] - d\psi^2,
\end{equation}
such that $dr^2 = dx^i \delta_{ij} dx^j$. The relevant nonzero
connections are
\begin{equation}
\Gamma^0_{ii} = - \frac{1}{\psi_0} e^{-2\psi^{-1}_0 t}, \quad
\Gamma^{\alpha}_{\alpha 4} = \frac{1}{\psi}, \quad \Gamma^{i}_{i
0}=-\frac{1}{\psi_0}, \quad \Gamma^4_{00} = \frac{\psi}{\psi^2_0}.
\end{equation}

Since the metric (\ref{m1}) is Riemann-flat (and therefore
Ricci-flat), hence it is suitable to describe a 5D vacuum
($G_{ab}=0$) in the framework of the IMT of gravity. With this aim
we shall consider the 5D action
\begin{equation}\label{act}
{\cal I} = {\Large\int} d^4 x \,  d\psi \sqrt{\left| g\right|}
\left( \frac{{\cal R}}{16\pi G}+ \frac{1}{2} g^{ab} \varphi_{,a}
\varphi_{,b} \right),
\end{equation}
where $g$ is the determinant of the covariant metric tensor
$g_{ab}$: $ g=\left(\frac{\psi}{\psi_0}\right)^8 e^{-6\psi^{-1}_0
t}$.

\subsection{Effective 4D dynamics of $\varphi$}

The effective 4D spacetime being described by the line element
\begin{equation}\label{met}
dS^2 = \left[\frac{\psi^2(t)}{\psi^2_0} - \dot\psi^2\right] dt^2
-\frac{\psi^2(t)}{\psi^2_0} \, e^{-2\psi^{-1}_0 t} dR^2,
\end{equation}
where the dot denotes the derivative with respect to $t$ and
$\psi_0$ is some constant. In order to consider $t$ as a cosmic
time, one must require that
\begin{equation}
\frac{\psi^2(t)}{\psi^2_0} - \dot\psi^2 =1,
\end{equation}
so that the foliation is described by
\begin{equation}\label{ue}
\psi(t)= \psi_0\, \cosh{\left(t/\psi_0\right)}, \quad \rightarrow
\dot\psi(t) = \sinh{\left(t/\psi_0\right)}.
\end{equation}
Finally, the metric (\ref{met}), for a foliation (\ref{ue}) is
described by
\begin{equation}\label{m2}
dS^2 = dt^2 - \cosh{(t/\psi_0)}^2 \, e^{2\psi^{-1}_0 t}\, dR^2,
\end{equation}
which describes an 3D (flat) spatially isotropic universe which is
collapsing with a scale factor $a(t) =
\cosh{\left(t/\psi_0\right)} \, e^{-\psi^{-1}_0 t}$, a Hubble
parameter $H(t)={\dot{a}\over a}$ and a deceleration parameter
$q=-{\ddot{a} a\over \dot{a}^2}$ given by (for $H_0=1/\psi_0$)
\begin{eqnarray}
&& H(t) = H_0 \left[ \tanh{\left(H_0 t\right)}
-1\right], \\
&& q(t) = - \frac{2\,\cosh{\left(H_0 t\right)}}{\cosh{\left(H_0
t\right)}-\sinh{\left(H_0 t\right)}}.
\end{eqnarray}
Notice that $\dot{H} >0$ and $\left.a(t)\right|_{t\rightarrow
\infty} 1/2$, such that the asymptotic size of the universe is
finite. Furthermore the late time asymptotic derivative the Hubble
parameter and the deceleration parameter, are
\begin{eqnarray}
&& \left.\dot{H}(t)\right|_{t\rightarrow \infty} \rightarrow 0, \\
&& \left.{q}(t)\right|_{t\rightarrow \infty} \rightarrow -\infty,
\end{eqnarray}
which means that the universe describes a collapse with asymptotic
Minkowski spacetime.

\subsection{Einstein´s equations}

On the other hand, the relevant components of the Einstein tensor
in Cartesian coordinates, are
\begin{eqnarray}
\bar{G}^0_{\,\,0} & = & -\frac{3 H^2_0}{\cosh^2{(H_0 t)}}
\left[\cosh{(H_0 t)} -\sinh{(H_0 t)}\right]^2, \label{30} \\
\bar{G}^i_{\,\,j} & = & -\frac{H^2_0}{\cosh^2{(H_0 t)}}
\left[\cosh{(H_0 t)}-\sinh{(H_0 t)}\right] \left[ 5\cosh{(H_0 t)}
- \sinh{(H_0 t)}\right] \delta^i_{\,\,j}, \label{31}
\end{eqnarray}
so that, using the fact that the Einstein equations are
respectively $G^{0}_{\,\,0} = -8\pi G\,\rho $ and $G^x_{\,\,x} =
G^y_{\,\,y}= G^z_{\,\,z}= 8\pi G \, P$, we obtain the equation of
state for the universe
\begin{equation}\label{state}
\frac{P}{\rho} = \omega(t) = - \frac{1}{3} \frac{\left[ 5
\cosh{(H_0 t)} - \sinh{(H_0 t)} \right]}{\left[\cosh{(H_0 t)}
-\sinh{(H_0 t)}\right]}.
\end{equation}
Notice that $\omega$ always remains with negative values
$\omega(t) <-1$, and evolves from $-5/3$ to $-\infty$, for large
asymptotic times. The effective 4D scalar curvature
\begin{equation}
{\cal{\bar{R}}} = \frac{6 H^2_0}{\cosh^2{(H_0 t)}}
\left[\cosh{(H_0 t)}-\sinh{(H_0 t)}\right] \left[ 3\cosh{(H_0 t)}
- \sinh{(H_0 t)}\right],
\end{equation}
decreases with the time and has a null asymptotic value
$\left.\cal{\bar{R}}\right|_{t\rightarrow \infty} \rightarrow 0 $.

The expectation values for the energy density and the pressure,
written in terms of the scalar field
$\varphi(t,\vec{r},\psi(t))\equiv \bar{\varphi}(t,\vec{r})$, are
\begin{eqnarray}
\bar{\rho} = \left<0|\bar{T}^0_{\,\,0} |0\right> & = & \left<
\frac{\psi^2_0}{\psi^2(t)} \left[\frac{1}{2}\dot{{\varphi}}^2 + \frac{1}{2
a^2(t)} \left(\vec{\nabla} {\varphi}\right)^2\right]+ \frac{1}{2}
 \left(\frac{\partial{\varphi}}{\partial \psi}\right)^2 \right>_{\psi(t)} \nonumber \\
 & = & \left<
\frac{1}{2} \dot{\bar{\varphi}}^2 + \frac{1}{2 a^2(t)} \left(\vec{\nabla}
\bar{\varphi}\right)^2  \right>, \label{34} \\
\bar{P}  = -\left<0| \bar{T}^i_{\,\,j}|0\right> & = & -
\delta^i_{\,\,j} \left<\frac{\psi^2_0}{\psi^2(t)} \left[ \frac{1}{2}
\dot{{\varphi}}^2 - \frac{1}{6 a^2(t)} \left(\vec{\nabla}
{\varphi}\right)^2\right] - \frac{1}{2}
 \left(\frac{\partial{\varphi}}{\partial \psi}\right)^2 \right>_{\psi(t)} \nonumber \\
 & = & -\delta^i_{\,\,j} \left< \frac{1}{2} \dot{\bar{\varphi}}^2 - \frac{1}{6 a^2(t)} \left(\vec{\nabla}
\bar{\varphi}\right)^2 \right>. \label{35}
\end{eqnarray}
Here, the notation $\left<0| ... |0\right>$ denotes the quantum
expectation value calculated on a 4D vacuum state. Because we are
considering a spatially isotropic and homogeneous background, we
shall consider an averaging value with respect to a Gaussian
distribution on a Euclidean 3D volume.

\section{Field dynamics and vacuum}

The effective 4D equation of motion for $\bar{\varphi}$, is
\begin{equation}
\ddot{\bar{\varphi}} - \frac{e^{2 H_0 t}}{\cosh^2{(H_0 t)}}
\bar{\nabla}^2_r \bar{\varphi} + 3 H_0 \,\left[ \tanh{(H_0 t)}
-1\right] \dot{\bar{\varphi}} =0,
\end{equation}
which in the limit of $t\rightarrow \infty$ tends to an equation
of motion for a massless scalar field on a asymptotic Minkowski
spacetime: $\ddot{\bar{\varphi}}-\bar{\nabla}^2_r
\bar{\varphi}=0$.

Using the eqs. (\ref{30}) and (\ref{31}) joined with eqs.
(\ref{34}) and (\ref{35}), we obtain from the effective 4D
Einstein equations (\ref{7}) the following relevant
non-perturbative expressions for the expectation values of squared
scalar field $\bar\varphi$:
\begin{eqnarray} \left<
\left(\vec{\nabla}\bar{\varphi}\right)^2 \right> &= & -\frac{3
H^2_0}{4\pi G} e^{-2 H_0 t} \left[ \cosh^2{(H_0 t)} - \sinh^2{(H_0
t)} \right], \label{36} \\
\label{a1} \left< \left(\dot{\bar{\varphi}}\right)^2 \right> & = &
- \frac{3 H^2_0 }{2\pi G} \left[ 1 - \tanh{(H_0 t)} \right],
\label{37}
\end{eqnarray}
that has the asymptotic large time limits
\begin{eqnarray}
\left< \left(\vec{\nabla}\bar{\varphi}\right)^2
\right>_{t\rightarrow \infty} &\rightarrow & 0 , \label{b1} \\
\left< \left(\dot{\bar{\varphi}}\right)^2 \right>_{t\rightarrow
\infty} &\rightarrow & 0.
\end{eqnarray}

\subsection{The modes}

If we redefine the modes $\chi_k(t) = a^{3/2} \xi_k(t)$, we obtain
the equation of motion for $\chi_k(t)$
\begin{equation}\label{chi}
\ddot{\chi}_k(t) + \left[ \frac{k^2}{a^2} - \left(\frac{9}{4}
H^2(t) + \frac{3}{2} \dot{H}(t)\right)\right]  \chi_k(t)=0,
\end{equation}
which has the general solution
\begin{eqnarray}
\chi_k(t) &=& e^{ 3 H_0 t} e^{\frac{1}{2} \ln{(1+ e^{2H_0
t})}}\left\{ A_k \,\,e^{-\frac{1}{2} \sqrt{1-(k/H_0)^2}
\ln{(1+e^{2H_0 t})}} \,\,_2F_1\left[ [a_1,b_1],[c_1]; 1+e^{2H_0 t}
\right] \right. \nonumber
\\
& + & \left. B_k \,\, e^{\frac{1}{2} \sqrt{1-(k/H_0)^2}
\ln{(1+e^{2H_0 t})}} \,\,_2F_1\left[ [a_2,b_2],[c_2]; 1+e^{2H_0
t}\right] \right\},
\end{eqnarray}
such that $_2F_1\left[ [a,b],[c]; x(t)\right]$ is the Gaussian
hypergeometric function with argument $x(t)= 1+e^{2H_0 t}$, and
\begin{eqnarray}
a_1 = \left. 2+ i \frac{k}{H_0} - \sqrt{1- \left(k/H_0\right)^2}
\right|_{k\gg H_0} \simeq 2, \\
b_1 = \left. 2- i \frac{k}{H_0} - \sqrt{1- \left(k/H_0\right)^2}
\right|_{k\gg H_0} \simeq 2\left( 1 - i \frac{k}{H_0} \right), \\
c_1 = \left. 1 - 2\sqrt{1- \left(k/H_0\right)^2} \right|_{k\gg
H_0} \simeq 1 - 2 i \frac{k}{H_0}, \\
a_2 = \left. 2+ i \frac{k}{H_0} + \sqrt{1- \left(k/H_0\right)^2}
\right|_{k\gg H_0} \simeq 2\left( 1 + i \frac{k}{H_0} \right), \\
b_2 =\left. 2- i \frac{k}{H_0} + \sqrt{1- \left(k/H_0\right)^2}
\right|_{k\gg H_0} \simeq 2, \\
c_2 = \left. 1 + 2\sqrt{1- \left(k/H_0\right)^2} \right|_{k\gg
H_0} \simeq 1 + 2 i \frac{k}{H_0}.
\end{eqnarray}
In this UV limit the Hypergeometric functions take the asymptotic
expressions
\begin{eqnarray}
\left._2F_1\left[[a_1,b_1],[c_1];e^{2H_0 t}\right]\right|_{UV}  &
\simeq & \frac{\left[1+
2 i k/H_0\right]}{\left[1- 2 i k/H_0\right]} e^{-4H_0 t}, \\
\left._2F_1\left[[a_2,b_2],[c_2];e^{2H_0 t}\right]\right|_{UV}  &
\simeq & \frac{\left[1- 2 i k/H_0\right]}{\left[1+ 2 i
k/H_0\right]} e^{-4H_0 t},
\end{eqnarray}
so that the large times asymptotic UV redefined modes $\chi_k(t)$
can be written as
\begin{equation}
\left.\chi_k(t)\right|_{H_0 t\gg 1, k/H_0 \gg 1} \simeq
\frac{A_k\, e^{-i k t} \left[1+ 2 i k/H_0 \right]^2 + B_k \, e^{ik
t} \left[1-2i k/H_0\right]^2}{ \left[ 1+ 4 \left(k/H_0\right)^2
\right]},
\end{equation}
where $A_k$ and $B_k$ are constants to be determined by
normalization of the modes on the effective 4D hypersurface:
$\chi_k \dot\chi^*_k - \dot\chi_k \chi^*_k = i$.  From this
condition we obtain that $\chi_k \chi^*_k = {1\over 2 k}$, so that
if we choose $B_k=0$, we obtain that $$A_k=-{1\over\sqrt{2 k}}
{\left[1+4(k/H_0)^2\right]\over \left[1+2 i k/H_0\right]^2},$$ and
the normalized modes $\chi_k$ on the UV sector become
\begin{equation}
\left.\chi_k(t)\right|_{H_0 t\gg 1, k/H_0 \gg 1} \simeq
-\frac{1}{\sqrt{2k}} e^{-i k t}.
\end{equation}
The final solution for the redefined modes are
\begin{equation}
\chi_k(t) = - {e^{ 3 H_0 t} e^{\frac{1}{2}
\left[1+\sqrt{1-(k/H_0)^2}\right] \ln{(1+e^{2H_0 t})}}
\left[1+4(k/H_0)^2\right]\over\sqrt{2 k} \left[1+2 i
k/H_0\right]^2}  \,\, \,\,_2F_1\left[ [a_2,b_2],[c_2]; 1+e^{2H_0
t}\right].
\end{equation}
Using the fact that $\left<
\left(\vec{\nabla}\bar{\varphi}\right)^2 \right> = - {1\over 2
\pi^2\, a^{3}} \int^{k_0(t)}_{0} k^4 \left(\chi_k \chi^*_k\right)
\, dk$, with the expression (\ref{36}), we obtain that the maximum
time-dependent wave-number, $k_0(t)$, is
\begin{equation}
k_0(t) = \left[ 12 \pi H^2_0 M^2_p \right]^{1/4} \, e^{-{5H_0\over
4} t}\, \cosh^{3/4}{(H_0 t)}\,\left[ \cosh^2{(H_0 t)} -
\sinh^2{(H_0 t)} \right]^{1/4},
\end{equation}
which tends to zero as $t\rightarrow \infty$.

\subsection{Classical and quantum contributions}

We consider the background solution of the field, which is given
by the zero mode ($k=0$) solution  $\chi_0(t)$ of the differential
equation (\ref{chi}). A particular solution for this equation is
\begin{equation}
\chi_0(t) = \frac{C}{\left[1+\tanh{(H_0 t)}\right]^{3/2}},
\end{equation}
such that $\xi_0(t) = a^{-3/2} \chi_0(t)$ comply with
$\dot\xi_0(t)=0$. Such that solution provide us the background
(classical) contribution for the field $\bar\varphi$. Therefore,
the results (\ref{36}) and (\ref{37}), for $\left<
\left(\vec{\nabla}\bar{\varphi}\right)^2 \right>$ and $\left<
\left(\dot{\bar{\varphi}}\right)^2 \right>$, respectively, has the
unique origin in the quantum contributions of the field. This
result is valid along all the collapse and has been calculated
exactly without using any approximation.

\section{Final Comments}

The idea that our universe is a 4D space-time embedded in a higher
dimensional has been a topic of increased interest in several
branches of physics, and in particular, in cosmology. This idea
has generated a new kind of cosmological models that includes
quintessential expansion. In particular, theories on which is
considered only one extra dimension have become quite popular in
the scientific community. Among these theories are counted the
braneworld scenarios \cite{bw}, the Induced Matter (IM) theory
\cite{STM,stm1,stm2,stm3,stm4} and all noncompact Kaluza-Klein
theories. The approach here considered is inspired in the IM,
where 4D sources appear as induced by one extended extra
dimension, meaning, by extended, that the fifth dimension is
considered noncompact. However, we have studied the case where the
foliation on the fifth extra coordinate is dynamical, so that the
resultant effective 4D scalar field $\bar\varphi(x^{\alpha})$ that
describes the collapsing system on the effective 4D hypersurface
(\ref{m2}) is dispersal. The important result here obtained relies
in that the origin of $\left<
\left(\vec{\nabla}\bar{\varphi}\right)^2 \right>$ and $\left<
\left(\dot{\bar{\varphi}}\right)^2 \right>$, is in the quantum
contributions of the field. This result is valid along all the
collapse and has been calculated exactly without using any
approximation.

\section*{Acknowledgements}

\noindent M.B. acknowledges UNMdP and CONICET Argentina for
financial support.

\end{document}